\documentclass{svjour2}                    
\smartqed  
\usepackage{graphicx}
\begin{document}

\title{The cyclotron gas stopper project at the NSCL
}



\author{C. Gu\'enaut         \and
        G. Bollen   \and S.\,Chouhan
     F.\,Marti \and D.J.\,Morrissey \and D.\,Lawton \and J.\,Ottarson \and G.K.\,Pang \and S.\,Schwarz \and B.M.\,Sherrill \and M.\,Wada \and A.F.\,Zeller }


\institute{C. Gu\'enaut \and G. Bollen \and S.\,Chouhan \and
F.\,Marti \and D.J.\,Morrissey \and D.\,Lawton \and J.\,Ottarson
\and G.K.\,Pang \and S.\,Schwarz \and B.M.\,Sherrill \and
A.F.\,Zeller \at
              Michigan State University, 1 Cyclotron, East Lansing MI-48824-1321, USA \\
              Tel.: +1-517-333-6463\\
              Fax: +1-517-353-5967\\
              \email{guenaut@nscl.msu.edu}           
           \and
           M. Wada \at
             RIKEN, 2-1 Hirosawa, Wako, Saitama, Japan
}

\date{Received: date / Accepted: date}

\maketitle

\begin{abstract}
Gas stopping is becoming the method of choice for converting beams
of rare isotopes obtained via projectile fragmentation and
in-flight separation into low-energy beams. These beams allow
ISOL-type experiments, such as mass measurements with traps or
laser spectroscopy, to be performed with projectile fragmentation
products. Current gas stopper systems for high-energy beams are
based on linear gas cells filled with 0.1-1 bar of helium. While
already used successfully for experiments, it was found that space
charge effects induced by the ionization of the helium atoms
during the stopping process pose a limit on the maximum beam rate
that can be used. Furthermore, the extraction time of stopped ions
from these devices can exceed 100\,ms causing substantial decay
losses for very short-lived isotopes. To avoid these limitations,
a new type of gas stopper is being developed at the NSCL/MSU. The
new system is based on a cyclotron-type magnet with a stopping
chamber filled with Helium buffer gas at low pressure. RF-guiding
techniques are used to extract the ions. The space charge effects
are considerably reduced by the large volume and due to a
separation between the stopping region and the region of highest
ionization. Cyclotron gas stopper systems of different sizes and
with different magnetic field strengths and field shapes are
presently investigated.

\end{abstract}

\section{Introduction}
\label{intro} Rare isotope production via relativistic projectile
fragmentation and in-flight separation produces nuclides with
short half-lives without limitations due to element selectivity.
Several fragmentation facilities exist worldwide. At all of them
slowing down and stopping of fast rare isotope beams is under
development. The goal is to produce low-energy beams that can then
be used for ISOL-type experiments, $i.e.$ experiments using
low-energy radioactive beams with small phase space volume, such
as mass measurements with traps or laser spectroscopy, or
post-accelerated for low-energy reaction studies.
LEBIT\,\cite{Rin06}, installed at the NSCL at MSU, was the first
to demonstrate that relativistic (150\,MeV/u) rare isotope beams
can be thermalized to low-energy ($\sim 5$\,keV)and be used for
precision experiments. A number of high-precision mass
measurements have already been performed, for example on
$^{38}$Ca\,\cite{Bol06}, $^{66}$As, $^{64}$Ge, $^{69}$Se, and
$^{40,42}$S. At GSI, a stopping test with a large linear gas
cell\,\cite{Tri04} was successfully carried out. At RIKEN, a laser
spectroscopy experiment with trapped radioactive Be ions, obtained
via gas stopping of fast fragments, was performed
recently\,\cite{Wad06}.

Present gas stopping schemes are all based on the slowing down of
the fast fragments in solid degraders and a final stopping of the
ions in a chamber filled with helium gas. Remaining singly or
doubly-charged, ions are guided out of the gas using electric
fields and gas flow and then prepared into a low-emittance,
low-energy ion-beam by means of radio-frequency (RF) ion guiding
techniques. Different concepts are applied for the ion extraction.
In the case of low-pressure gas cells ($< 300$\, mbar He) a
combination of electrostatic and RF potentials is often employed.
The gas cell\,\cite{Wei04} of LEBIT is operated at high pressure
(1\,bar He). Static electric fields guide the ions to an
extraction nozzle where the force provided by the gas flow
transports them out of the gas cell.

A rate-dependent efficiency for linear gas cells has been observed
in a variety of systems\,\cite{Wad03,Wei05,Tak05,Fac06}.
Extraction efficiencies of existing linear gas cells decrease
precipitously with the ionization rate density (rate of generation
of ion pairs (IP) per volume) inside the gas cell. The decrease in
efficiency is attributed to space-charge effects, which lead to
ion losses inside the gas cell. Next generation facilities will
offer exotic beam rates of $10^9$\,/s, requiring the efficient
handling of ionization rate densities of about $10^{11}$
IP/cm$^3$/s. This is not achievable with existing gas cells
without a significant loss in efficiency.

For the stopping of rare isotope beams with an energy of about
100\,MeV/u, linear gas cells need to be operated with a
pressure-length product of typically $p \cdot L$=0.5 bar$\cdot$m.
Limited by the maximum applicable electric field for ion transport
and extraction inside the gas cell the average extraction time is
about 100\,ms. Such long extraction times do not match the
advantage of fast-beam production and lead to decay losses.

In order to maximize the benefit of the gas stopping approach, the
following requirements have to be fulfilled:

\begin{itemize}
    \item Short extraction times. In order to minimize decay losses the
extraction time should be comparable or shorter than the shortest
half-life of the ions to be studied. Extraction times of 10 ms or
less are desirable.
    \item Efficient stopping and extraction at high beam
    intensities. Next-generation facilities will provide
rare isotope beam intensities of up to $10^9$\,/s, many orders of
magnitude higher than available at present fragmentation
facilities.
    \item Applicability to all fragment beams. In order to be universal,
the gas stopper needs to be able to handle beams of isotopes with
largely different charges $Z$ and neutron-to-proton ratios.
\end{itemize}

A new concept, the cyclotron gas stopper, promises to fulfill
these requirements and to overcome the limitations of linear gas
stoppers \cite{Bol05}. Such a system, based on a cyclotron-type
focusing magnet with a gas-filled stopping chamber and using
radio-frequency (RF) ion guiding techniques for ion extraction, is
presently under development at the NSCL.

\section{Concept}
\label{concept}

\begin{figure}
\includegraphics[width=0.75\textwidth]{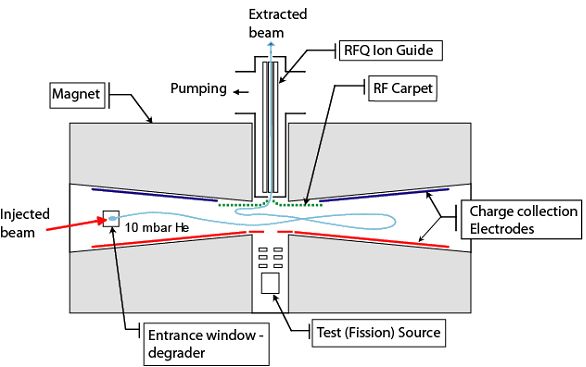} %
\caption{Schematic view of the cyclotron gas stopper showing the
main components of the system. A cyclotron-type focusing magnet
contains a  vacuum chamber filled with helium at low pressure, a
beam degrader, charge collection electrodes, and an RF carpet for
ion extraction.} \label{fig:concept}
\end{figure}

Figure\,\ref{fig:concept} presents the main components of the
cyclotron gas stopper. Ions injected into the system will first
interact with a solid degrader and then be slowed down in helium
gas at low pressure. The focusing properties of the magnet confine
the ions during the deceleration process. The ions are finally
extracted by means of static electric fields, an RF
carpet\,\cite{Wad03} and radio-frequency ion guides.

With a long stopping path, a low pressure may be used inside the
cyclotron gas stopper. This low pressure will allow for a fast
drift inside the magnet and a fast extraction, which will match
the advantages of fast-beam production. The larger volume as
compared to linear gas cells and a separation of the stopping
region from the primary ionization directly contribute to the
minimization of space charge effects.

A similar concept has been used for the production of
antiprotonic, pionic and muonic atoms\,\cite{Sim93} and has also
been discussed for the stopping of light ions\,\cite{Kat98}. The
benefit of this concept for the stopping of intense rare isotope
beams was first shown in simulations performed at the
NSCL/MSU\,\cite{Bol05}.

An RF carpet has already been used successfully\,\cite{Wad03} in a
linear gas cell. Because of its low pressure ($<20$ mbar) the
cyclotron gas stopper provides an ideal environment for the
operation of RF carpets. The modest damping of the ion motion, as
compared to high-pressure linear gas cells, allows carpets to be
used with a relatively large pitch and low voltages, while still
providing a strong repelling force. Static potentials will be used
to guide the ions onto the carpet surface and to the extraction
orifice.

\section{Mechanical design} \label{design}

\begin{figure}
\includegraphics[angle=-90,width=0.75\textwidth]{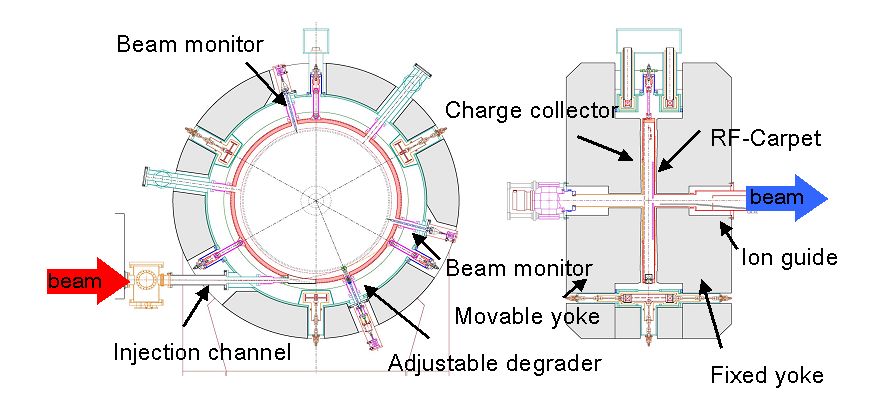} %
\caption{Design concept of a vertical superconducting magnet under
consideration.} \label{fig:design}
\end{figure}

The mechanical design of the cyclotron gas stopper is underway and
will be based on detailed simulations discussed below. At the
present stage of design a vertically oriented magnet system is
favored. Superferric magnets have been designed that produce peak
magnetic fields between 1.6 and 3\,T. Two separated coil packages
will be used in order to be able to open the yoke, facilitating
the access to the inner part of the system.
Figure\,\ref{fig:design} presents a conceptual design of one of
the systems under consideration. The vertical arrangement has
advantages in particular for the extraction and transport of the
low-energy beams out of the fringe field. The magnet system will
house a cryogenically cooled vacuum chamber, filled with helium
gas at a typical pressure of 10\,mbar or less. The beam degrader
and beam monitors are inserted radially. The diameters of the
systems presently considered range from $3-4$\,m with beam
injection radii between $0.7-1.5$\,m. Options considered are
sector fields for stronger transverse focusing, different field
shapes, and the use of multiple degraders.

\section{Beam stopping simulations} \label{simu}

Various detailed numerical calculations based on realistic
magnetic fields are being carried out to optimize and characterize
the system. They include the Lorentz force, energy loss,
charge-exchange collisions, and small-angle
multiple-scattering\,\cite{Sig74}. The simulations are being
performed for light to heavy isotopes of key nuclides with
different $A/Z$. Bromine isotopes $^{70,79,94}$Br were chosen to
represent the central region of the nuclear chart. Iodine isotopes
$^{108,127,144}$I for heavier isotopes and $^{6,9,11}$Li for the
very lighter mass region. The choice of these nuclides was also
based on the availability of data for low-energy charge-exchange
cross sections with helium\,\cite{Bet72,Woi92}.

A C++ version\,\cite{Lise} of the ATIMA\,\cite{Atima} code is used
to model the interaction of the incoming beam with the degrader.
ATIMA calculates the stopping power, the energy loss, the
energy-loss straggling, the angular straggling, the range, and the
range straggling.

\begin{figure}
\includegraphics[width=0.75\textwidth]{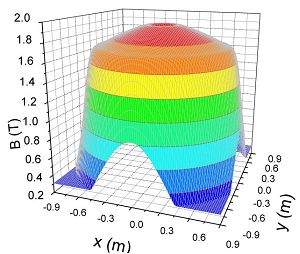} %
\caption{Realistic magnetic fields obtained in a weakly focusing
magnet (left) and a sector-field magnet (right), calculated with
the TOSCA code.} \label{fig:magnet}
\end{figure}

Different magnet sizes and field shapes are under investigation.
Weakly focusing and a sector-field magnet are considered. Examples
for the resulting fields are shown in Figure\,\ref{fig:magnet}.
Both systems allow ions to be stopped effectively. The
sector-field magnet offers some advantages since the injection for
isotopes with large $A/Z$ (high-rigidity) is simplified.

\begin{figure}
\includegraphics[width=11cm]{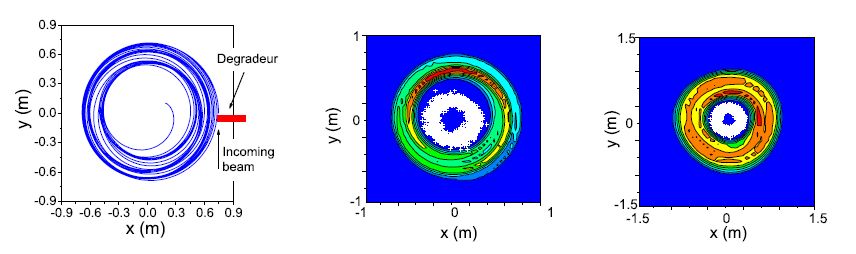} 
\caption{Trajectory of a single $^{79}$Br ion with an energy of
610\,MeV after the degrader inside the weakly focusing magnet
filled with 10\,mbar He (left). Energy deposition and stopped ion
distribution (crosses) for a weakly focusing magnet (middle) and a
sector-field magnet (right).} \label{fig:energy_loss}
\end{figure}

Beam simulations were for example performed for $^{79}$Br isotopes
with 100\,MeV/u before the degrader in a weakly focusing magnet
with $B_{max}=2$\,T and with 60\,MeV/u in the case of the
sector-field magnet with $B_{max}=2.6$\,T. In both cases helium
gas pressure of about 10\,mbar was used.
Figure\,\ref{fig:energy_loss} shows typical results. On the left
the trajectory of a single ion inside the weakly focusing magnet
is presented. The two figures on the right show stopped ion
distributions (crosses) together with energy loss (ionization)
densities (colored/greyscale areas) inside the gas. A key
advantage of the cyclotron gas stopper is that there is a
separation in-between the region of highest ionization and the
central region were the ions stop. Compared to linear stoppers
this leads to a large reduction of space charge effects.

Stopping and extraction efficiencies higher than 95\,\% have been
achieved for the bromine beams so far investigated.
The systematic exploration of the stopping properties for other
rare isotope beams is on its way.

%

%

\begin{acknowledgements}
This work has been supported by MSU and by DOE under contract
number DE-FG02-06ER41413.
\end{acknowledgements}



\end{document}